\documentclass[letterpaper, 10 pt, conference]{ieeeconf}  

\IEEEoverridecommandlockouts                              

\usepackage{graphicx}

\title{\LARGE \bf
Upper Limb Movement Execution Classification using Electroencephalography for Brain Computer Interface
}

\author{Saadat Ullah Khan$^{1}$, Muhammad Majid$^{1}$, Syed Muhammad Anwar $^{2,3}$ 
\thanks{*This work was not supported by any organization}
\thanks{$^{1}$ Saadat Ullah Khan and Muhammad Majid are with Department of Computer Engineering, University of Engineering and Technology, Taxila, Pakistan.}%
\thanks{$^{2}$ Syed Muhammad Anwar is with Sheikh Zayed Institute for Pediatric Surgical Innovation, Children’s National Hospital, Washington, DC.}%
\thanks{$^{3}$ Syed Muhammad Anwar is with School of Medicine and Health Sciences, George Washington University, Washington, DC.}
}

\begin{document}

\maketitle
\thispagestyle{empty}
\pagestyle{empty}

\begin{abstract}
An accurate classification of upper limb movements using electroencephalography (EEG) signals is gaining significant importance in recent years due to the prevalence of brain-computer interfaces. The upper limbs in the human body are crucial since different skeletal segments combine to make a range of motion that helps us in our trivial daily tasks. Decoding EEG-based upper limb movements can be of great help to people with spinal cord injury (SCI) or other neuro-muscular diseases such as amyotrophic lateral sclerosis (ALS), primary lateral sclerosis, and periodic paralysis. This can manifest in a loss of sensory and motor function, which could make a person reliant on others to provide care in day-to-day activities. We can detect and classify upper limb movement activities, whether they be executed or imagined using an EEG-based brain-computer interface (BCI). Toward this goal, we focus our attention on decoding movement execution (ME) of the upper limb in this study. For this purpose, we utilize a publicly available EEG dataset that contains EEG signal recordings from fifteen subjects acquired using a 61-channel EEG device. We propose a method to classify four ME classes for different subjects using spectrograms of the EEG data through pre-trained deep learning (DL) models. Our proposed method of using EEG spectrograms for the classification of ME has shown significant results, where the highest average classification accuracy (for four ME classes) obtained is 87.36\%, with one subject achieving the best classification accuracy of 97.03\%. \newline

\indent \textit{Clinical relevance}— This research shows that movement execution of upper limbs is classified with significant accuracy by employing a spectrogram of the EEG signals and a pre-trained deep learning model which is fine-tuned for the downstream task.
\end{abstract}

\section{INTRODUCTION}
A brain-computer interface (BCI) aims to provide a channel of communication between the human brain and an external system for its control [1]. Non-invasive BCI using electroencephalography (EEG) is beneficial since it provides the ability to acquire and decode the underlying electrical activity of the brain from the scalp using EEG electrodes [2]. With the commercial availability of EEG recording devices and publicly available datasets, it is becoming a mature topic of research. Non-invasive EEG has many advantages compared to other modalities that have been explored for the restoration of upper limb movements in the past. For instance, it does not require surgical intervention, it is painless, cheap, portable, and accurate [3]. Hence EEG is widely used in various biomedical applications including seizure detection [4], stress assessment [5], depression disorder detection [6], and schizophrenia [7]. 
The interaction of the human body with its environment depends significantly on controlled movements of the upper limbs. However, spinal cord injury (SCI) and other neuro-muscular diseases can effect this control and limb movement. 
It is crucial to restoring upper limb movements for people with SCI so they can independently take care of their daily activities. It has been found that movement execution (ME) generates stronger amplitude co-relates in EEG signals, hence decoding ME gave promising results when compared to decoding movement imagination (MI) [8]. 

\begin{figure*}[t]
  \centering
  \centerline{
  \includegraphics[width = 150mm]{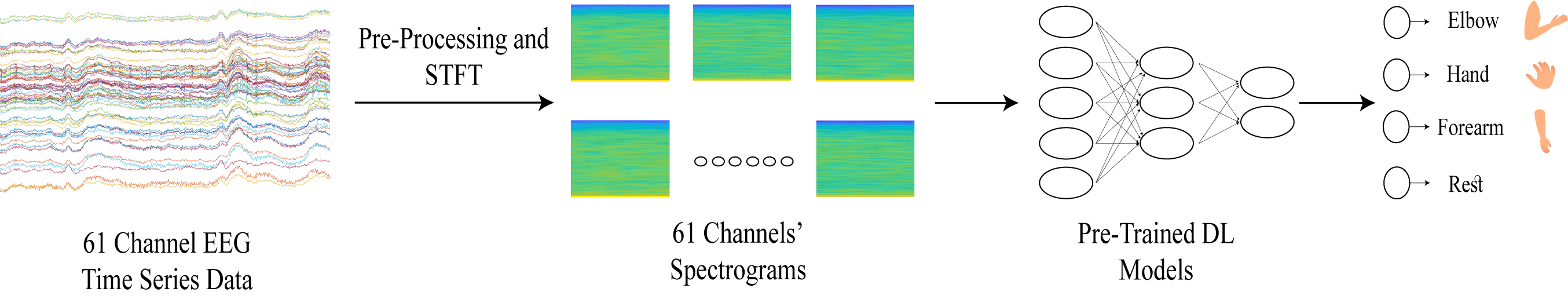}}
  \caption{Our proposed methodology for motion execution (ME) classification using EEG spectrograms.}\medskip
  \label{fig:fig1}
\end{figure*}

Deep learning (DL) methods are getting a lot of attention recently in a diverse set of data processing tasks. The same holds for upper limb movement classification using EEG data. 
Features like minimal pre-processing of data, automatic low- and high-level feature extraction, and superior learning capabilities make DL an attractive tool in the machine learning domain [9]. 
Classification of the MI of two movements was performed using a convolutional neural network (CNN),  along with common spatial patterns (CSP) as a feature extraction mechanism beforehand. The strategy was able to classify palm extension and hand grasp movements [10]. Using a CNN consisting of two convolution layers followed by fully connected (FC) layers that use a two-dimensional representation of the input EEG signal, 
was proposed for the classification of grasping movements in the upper limb [11]. An encoder-decoder architecture was proposed for the classification of four MI classes using EEG data, where the encoder was used for efficient representation of the input EEG signals and the decoder for the classification of EEG data into four classes [12]. Classification of both ME and MI using EEG was performed using shrinkage regularized linear discriminant analysis (sLDA). Using a time window and a single point in time of EEG signals, sLDA has shown the ability to perform binary classification of EEG signals [13]. Time series EEG signals have recently been used with gated transformers, which possess the ability to capture long-term dependencies, for the classification of EEG data. The classification accuracy of the model varies depending on the time window of the EEG data used. If the time window of the EEG signal is wide then the model achieves acceptable performance [14]. By introducing an attention mechanism to a long short-term memory network, binary classification of right and left-hand movement was performed [15]. Using the spatial and temporal information of the EEG data the raw time-series signals can be converted into a three-dimensional input. Using a multi-branch CNN model that uses this three-dimensional EEG data has shown the capability of classifying certain movement intentions of a few movement states [16]. A hierarchical DL model was proposed consisting of three CNNs that used feature sharing for the classification of up to nine classes of MI using EEG data [17]. In another study, a one-versus-rest classification strategy with CNN was proposed, where one movement class was classified against all other movement classes. Using filter bank common spatial patterns (FBCSP) as a feature extractor with a CNN consisting of three convolution layers, the classification of three individual movements was performed [18]. Binary classification of hand movements like raising either the right or the left arm was performed successfully by using principal component analysis as feature extraction and dimensionality reduction algorithm and a three-layered CNN model [19].

In summary, the utilization of spectrograms from EEG signals for ME classification tasks is an area not explored. Herein to the best of our knowledge, for the first time we propose to perform the classification of ME tasks using the spectrogram of the EEG data. Particularly, EEG data has been used for upper limb movement detection and, since ME tasks are easier to perform, these EEG signals would be more consistent [8]. Using appropriate spectrograms, we can better encode the time and frequency information within the EEG signal. 
For the classification of the spectrograms, we use pre-trained DL models. Further, we experimented with DL models of different depths to understand the effect on classification performance. We use intra-subject training for the classification of the ME spectrograms i.e., the model is trained according to the number of subjects.


\section{Proposed Methodology}
Our proposed methodology aims to classify four classes of ME based on the spectrograms of the time-series EEG data. To achieve this, we compute a spectrogram for data from each EEG channel and use it as input to the DL models. The four classes include hand movement, forearm movement, elbow movement, and rest. A block diagram of the proposed methodology for ME classification from EEG spectrograms for BCI is shown in Fig. \ref{fig:fig1}.

\subsection{EEG Dataset}
We used upper limb movement decoding from EEG (001-2017) dataset [13] for the classification of ME. This dataset is publicly available and consists of seven classes for upper limb movement and EEG recordings of $15$ healthy subjects with ages ranging from 22 to 40 years. The EEG data were acquired using a 61-channel EEG headset with a sampling frequency of 512 Hz. The seven movements in this dataset are elbow extension, elbow flexion, hand close, hand open, forearm supination, forearm pronation, and rest. For our methodology, we convert this seven-class EEG dataset into four classes by merging common classes i.e., hand open and hand close classes to one hand movement class, elbow extension and elbow flexion classes into elbow movement class, and finally, the forearm supination and forearm pronation classes to forearm movement class.


\begin{figure}[t]
  \centering
  \centerline{
  \includegraphics[width = 80mm]{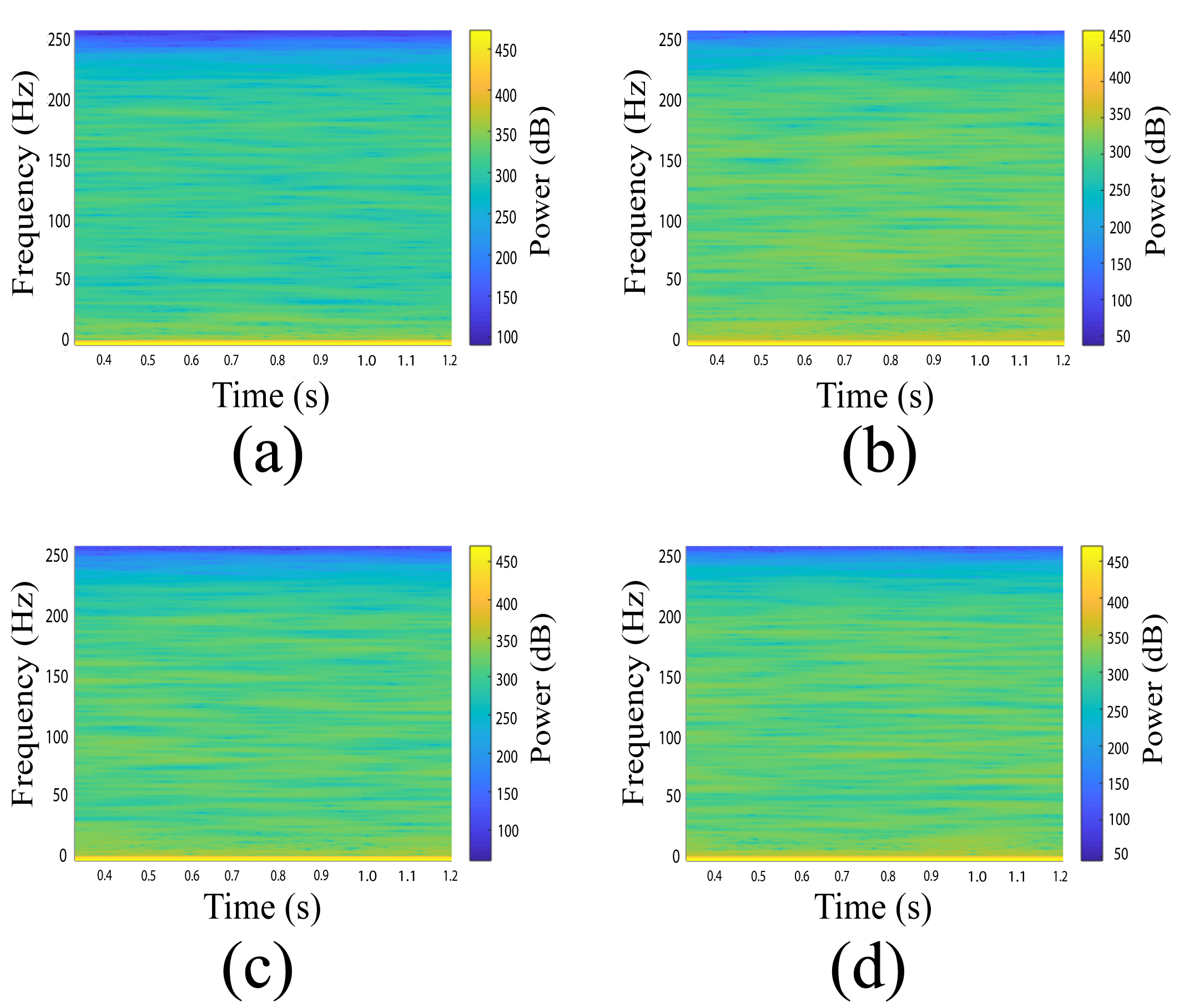}}
  \caption{Spectrograms of EEG data for different ME (a) elbow movement, (b) hand movement, (c) forearm movement, (d) rest.}\medskip
  \label{fig:fig2}
\end{figure}

\subsection{EEG Pre-processing and Spectrogram Generation}
To remove noise from the time-series EEG data, an $8^{th}$ order Chebyshev filter was applied ranging from 0.01 Hz to 200 Hz. To eliminate power line interference from the EEG data, a 50 Hz notch filter was applied. To compute the spectrogram of the EEG data we use a short-time Fourier transform (STFT) given in Eq. \ref{eqn:stft} for each channel of the time-series EEG data.

\begin{equation}
\label{eqn:stft}
X_{m}(\omega) = \sum_{n=-\infty}^{\infty} x(n + mR) w(n)e^{-\iota\omega(n + mR),}
\end{equation}
where $X_{m} (\omega)$ is the short time Fourier transform (DTFT) of the input data, $x$ represents the input signal at $n$, $\omega(n)$ is the window used of length $m$, and $R$ is the number of samples between successive DTFTs.


\subsection{Pre-Trained DL models}

For the classification task, we use pre-trained DL models implemented in PyTorch. These DL models are pre-trained using the ImageNet-1k dataset. Specifically, we use VGG for ME classification. To study the effect of the depth of the network we use VGG-11, VGG-13, VGG-16, and VGG-19. To achieve four class classifications of ME from the EEG spectrogram, we replace the final layer of the model with a fully connected layer (FCL) consisting of four neurons. The input dimensions of the DL models employed are $H \times W \times C$ with the number of components ($C$) required to be 3. For this purpose, we stack the values of a spectrogram into three components that make the single-component spectrogram into a multi-component spectrogram.

\begin{table}[t]
\caption{Classification accuracy of the proposed ME classification for BCI using EEG spectrograms and pre-trained DL model.}
\label{table_1}
\begin{center}
\begin{tabular}{|c||c||c||c||c|}
\hline
Subject & VGG-11 & VGG-13 & VGG-16 & VGG-19\\
\hline
S1 & 86.51\% & 90.88\% & 90.02\% & 91.19\%\\
\hline
S2 & 88.68\% & 78.53\% & 86.33\% & 59.34\%\\
\hline
S3 & 76.69\% & 79.97\% & 85.34\% & 67.27\%\\
\hline
S4 & 78.12\% & 75.33\% & 79.11\% & 82.74\%\\
\hline
S5 & 83.27\% & 83.56\% & 78.14\% & 85.55\%\\
\hline
S6 & 80.52\% & 86.01\% & 44.53\% & 86.88\%\\
\hline
S7 & 91.80\% & 91.70\% & 59.68\% & 87.90\%\\
\hline
S8 & 97.03\% & 96.01\% & 96.70\% & 96.68\%\\
\hline
S9 & 88.70\% & 84.75\% & 88.54\% & 88.72\%\\
\hline
S10 & 93.59\% & 93.91\% & 87.72\% & 95.39\%\\
\hline
S11 & 89.11\% & 66.82\% & 94.47\% & 91.37\%\\
\hline
S12 & 90.00\% & 86.06\% & 90.67\% & 87.92\%\\
\hline
S13 & 84.07\% & 87.97\% & 82.94\% & 81.49\%\\
\hline
S14 & 87.39\% & 91.49\% & 84.13\% & 91.35\%\\
\hline
S15 & 94.96\% & 95.55\% & 92.74\% & 95.60\%\\
\hline
\textbf{Average} & \textbf{87.36}\% & 85.90\% & 82.73\% & 85.95\%\\
\hline

\end{tabular}
\end{center}
\end{table}

\begin{figure}[t]
  \centering
  \centerline{\includegraphics[width = 85mm]{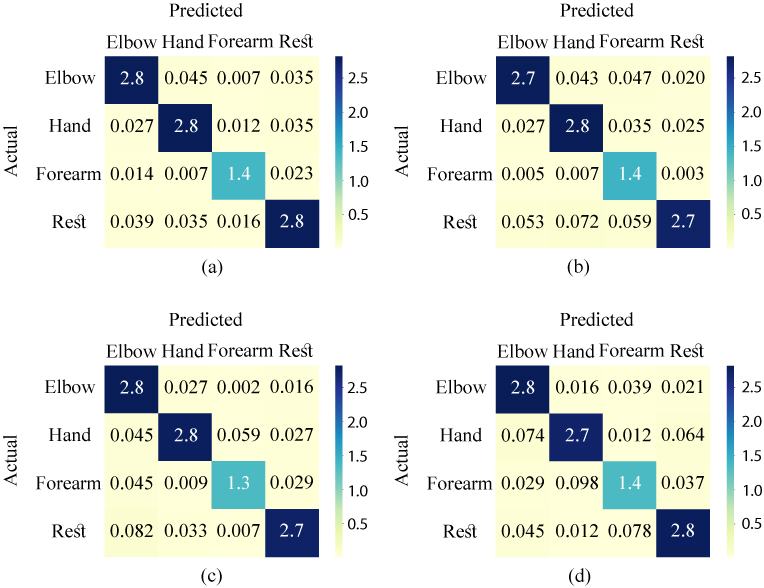}}
  \caption{Confusion matrices for subject 8: (a) VGG-11, (b) VGG-16, (c) VGG-16, (d) VGG-19.}\medskip
  \label{fig:fig3}
  
  \vspace{-7mm}
\end{figure}

\section{Results and Discussion}
The input dimension of the pre-trained DL models is $H \times W \times C$, with $H \& W$ at 224. Therefore, the input to the model is kept at the same dimensions. To compute a spectrogram of size $224 \times 224$ we select the first $788$ samples from the start of each trial of the EEG data. To achieve this specific height of the spectrogram we fix the number of discrete Fourier transform (DFT) points at $447$. To achieve the specific width of the spectrogram we fix $342$ and $340$ as the window size and the number of overlapping samples, respectively. The window used during the computation of the spectrogram was the Blackman window. To eliminate the imaginary values from the spectrogram we take the natural log of the absolute square of the values, and doing so results in a spectrogram with real values. Fig. \ref{fig:fig2} shows example spectrograms of the channel $C6$ for the four movements. We expect that the pictorial difference in the spectrograms would better enable the DL model to classify the data. Through our investigation, we have found that this specific width of the spectrogram can also be achieved through other combinations of the window size and the number of overlapping samples. These different cases of spectrograms could effect the output of the DL models employed. For this study, we have employed only one case of the spectrogram and left the effect of different cases of the spectrogram to future studies.

As we use an intra-subject training strategy for each subject, we fine-tuned the pre-trained DL models 15 times i.e., the model was trained individually for each subject and tested for that subject only. For this purpose, as there are $10$ EEG runs with $6$ trial of each class per run and $61$ total channels, we had a total of $7320$ spectrograms for each movement class except the rest class which had $3660$ number of training samples irrespective of the training-validation-test split. The number of spectrograms of the movement classes is double compared to the rest class because as we merge the two movement classes into one, the number of spectrograms also increases. We used 7:1:2 as the training-validation-testing ratio. A batch size of 16 was used, at 20 epochs for the fine-tuning of VGG. We observed an infinite/undefined loss at every epoch during the fine-tuning of the model. To mitigate this effect, we specifically use VGG with batch normalization. We kept the momentum at 0.9 and the learning rate at 0.001 throughout. For the loss function and the optimizer, we employed cross-entropy and stochastic gradient descent (SGD) respectively. This experimentation was performed on a PC with 64 gigabytes RAM, Intel\textregistered Xeon\textregistered W-2265 CPU @ 3.50GHz, and Nvidia\textregistered RTX \textsuperscript{TM} A5000 GPU.


Table \ref{table_1} shows the results of the four class ME spectrogram classification for the VGG models used. 
The average classification accuracy of the model is the highest for VGG-11. 
Whereas, VGG-19 achieves the second highest classification results with 85.95\% average classification accuracy. For each DL model used in this study, we observed that subject 8 achieves the highest classification accuracy. The confusion matrices for subject 8 for all models are shown in Fig. \ref{fig:fig3}. From the confusion matrices, it is evident that all models can't classify the forearm movements well. All models show similar true positive values, while the false negative and false positive values are less than the true positive and true negative values.

For a fair comparison of our proposed methodology of ME classification, we apply three-class classification on our data and compare our classification results with studies that use the same dataset. For this purpose, the three classes we classify are elbow movement, forearm movement, and hand movement. We do this by applying a linear layer with three neurons at the end of the DL model. Table \ref{table_2} shows the performance comparison of our methodology with what is reported in the literature. Our proposed methodology outperforms the FBCSP + CNN methodology in terms of the average classification achieved. Only subject $3$ performs worse in our methodology. 
We see a considerable improvement ($>$6\%) in performance using our proposed methodology for ME classification compared to what is reported in the literature.

\begin{table}[t]
\caption{Performance comparison of our proposed ME classification for three classes.}
\label{table_2}
\begin{center}
\begin{tabular}{|c|c|c|c|c|}
\hline
Subject & Proposed & Proposed & FBCSP+CNN \\
Number & VGG-11 & VGG-19 & [18]\\
\hline
S1 & 88.73\% & \textbf{92.53}\% & 82.57\%\\
\hline
S2 & 88.43\% & \textbf{89.02}\%  & 84.48\%\\
\hline
S3 & 82.76\% & 86.52\%  & 89.82\%\\
\hline
S4 & 84.01\% & \textbf{86.36}\%  & 84.66\%\\
\hline
S5 & 84.97\% & \textbf{87.72}\%  & 82.09\%\\
\hline
S6 & 85.08\% & \textbf{87.59}\%  & 87.30\%\\
\hline
S7 & 90.96\% & \textbf{92.91}\%  & 87.42\%\\
\hline
S8 & 97.40\% & \textbf{97.40}\%  & 80.52\%\\
\hline
S9 & 88.38\% & \textbf{88.59}\%  & 78.62\%\\
\hline
S10 & 95.10\% & \textbf{95.26}\%  & 83.67\%\\
\hline
S11 & 90.00\% & \textbf{96.96}\%  & 84.53\%\\
\hline
S12 & 90.09\% & \textbf{91.30}\%  & 82.98\%\\
\hline
S13 & \textbf{86.88}\% & 81.99\%  & 81.01\%\\
\hline
S14 & 90.25\% & \textbf{92.66}\%  & 84.11\%\\
\hline
S15 & 95.76\% & \textbf{96.03}\%  & 80.32\%\\
\hline
Average & 89.25\% & \textbf{90.25}\%  & 83.61\%\\
\hline

\end{tabular}
\end{center}
\end{table}

\section{CONCLUSION}
We successfully performed the classification of four upper limb movements using spectrograms of the ME EEG data through pre-trained DL models using a publicly available dataset. Classification of ME through spectrogram of the time series EEG has not been done before to the best of our knowledge. Particularly, we used the VGG DL model and also studied the effect on the classification accuracy as the depth of the DL model is increased. Through the application of DL on EEG spectrograms we found that VGG-11 performs the best among other VGG models achieving the highest classification accuracy of 97.03\% and average classification accuracy of 87.36\%. In the future, we intend to extend this work to a higher number of classes. 

\addtolength{\textheight}{-12cm}   






\end{document}